\documentstyle[12pt]{article}

\newcommand{\be}{\begin{equation}}
\newcommand{\ee}{\end{equation}}
\newcommand{\ba}{\begin{eqnarray}}
\newcommand{\ea}{\end{eqnarray}}
\newcommand{\ban}{\begin{eqnarray*}}
\newcommand{\ean}{\end{eqnarray*}}

\newcommand{\la}{\langle}
\newcommand{\ra}{\rangle}
\newcommand{\ket}[1]{| #1 \ra}
\newcommand{\bra}[1]{\la #1 |}

\newcommand{\braket}[1]{\la #1 \ra}

\newcommand{\bel}[1]{\begin{equation}\label{#1}}  
\newcommand{\beal}[1]{\begin{eqnarray}\label{#1}} 
\newcommand{\bmat}[1]{\lf(\begin{array}{#1}}	  
\newcommand{\emat}{\end{array}\r)}		  

\begin{document}

\title{The Bosonic Structure of Fermions\rule{0cm}{2cm}}
\author{J. Ruan\thanks{On leave from the Department of Modern Physics,
USTC, P.O. Box 4, Hefei, China; E-mail address:
jruan@physics.adelaide.edu.au}\hspace{1.6mm} and
R.J. Crewther\thanks{E-mail address:
rcrewthe@physics.adelaide.edu.au}\\
Department of Physics and Mathematical Physics\\
University of Adelaide\\
Adelaide, SA 5005\\
Australia}
\date{}
\maketitle
\begin{abstract}            
We bosonize fermions by identifying their occupation numbers as the
binary digits of a Bose occupation number. Unlike other schemes,
our method allows infinitely many fermionic oscillators to be
constructed from just one bosonic oscillator.
\end{abstract}
\vfill ADP-94-13/T154\\July 1994
\newpage
The idea that fermion operators can be constructed from bosonic
variables has a long history \cite{JW,Klaiber}. These days,
bosonization \cite{boson} is an essential tool for research in
field and string theory, especially in two space-time dimensions \cite{2d}.

A feature of existing schemes is that an infinite set of fermionic operators
$d_k$ and $d_{k}^{\dagger}$ labelled by an index or parameter $k$
is obtained by introducing a similar multiplicity of bosonic operators
$b_k$ and $b^\dagger_k$. This is particularly evident when $k$
takes discrete values, as in the model of Naka \cite{Naka}. Given that
each boson number operator $b^\dagger_k b_k$ can take any integer value
from 0 to $\infty$, whereas each fermion number operator $d^\dagger_k d_k$
has only two eigenvalues, 0 and 1, there would appear to be a lack of economy
in these procedures. It should be possible to produce an infinite
set of fermion oscillators from just {\em one\/} boson
oscillator\footnote{In \cite{chinese}, a {\em single\/} fermion oscillator
is constructed from one boson oscillator.}. We demonstrate that this can be
done.

Let $b^\dagger$ and $b$ be the creation and annihilation operators
of a single Bose oscillator:
\be
            bb^{\dagger}-b^{\dagger}b=1                        \label{one}
\ee
Then there is a number operator $N=b^{\dagger}b$ with the usual
algebraic properties
\be
b^{\dagger}N=(N-1)\, b^{\dagger}\ , \hspace{4mm} b\, N=(N+1)\, b  \label{two}
\ee
or more generally
\be
b^\dagger f(N) = f(N-1)\, b^\dagger\ , \hspace{4mm} b\, f(N) = f(N+1)\, b
\ee
for suitable functions $f$ of $N$. Our task is to construct operators
\[ d_k = d_k(b,b^\dagger)\ , \hspace{4mm}
   d^\dagger_k = d^\dagger_k(b,b^\dagger)  \]
for $k = 1,2,3, \ldots \infty\,$ such that the anticommutation relations
\ba\nonumber
\{d_{k},\;\;d_{k'}^{\dagger}\}&=&\delta_{kk'}  \nonumber    \\
\{d_{k},\;\;d_{k'}\}&=&0             \nonumber     \\
\{d_{k}^{\dagger},\;\;d_{k'}^{\dagger}\}&=&0
\label{anticom}
\ea
are satisfied.

Accordingly, we consider the vector space spanned by eigenstates
$\ket{n}$ of $N$, with each eigenstate normalized to unity:
\be
N\ket{n} = n\ket{n}\ , \hspace{4mm}\braket{m|n} = \delta_{mn}\ ,
\hspace{4mm}m,n = 0,1,2, \ldots \label{eigen}
\ee
Then there are standard formulas for the boson annihilation operator $b$,
creation operator $b^{\dagger}$ and number operator $N$,\newpage
\ba\nonumber
b&=&\sum_{n=0}^{\infty}\sqrt{n+1}\;\ket{n}\bra{n+1} \nonumber \\
b^{\dagger}&=&\sum_{n=0}^{\infty}\sqrt{n+1}\;\ket{n+1}\bra{n} \nonumber  \\
N&=&\sum_{n=0}^{\infty}n\;\ket{n}\bra{n}              \label{boseops}
\ea
and for the action of powers of $b$ and $b^\dagger$ on the ground state
$\ket{0}$:
\be
\ket{n}=\frac{1}{\sqrt{n!}}\ (b^{\dagger})^n \ket{0}\ ,\hspace{4mm}
0 = b\ket{0}        \label{seven}
\ee
We economise our notation by using the same symbol for an operator
\be  q = \sum_{mn} q_{mn} \ket{m} \bra{n}  \label{op} \ee
and its matrix representative $q = (q_{mn})$, writing e.g.
\[
b=\left(\begin{array}{cccccc}
    0 & \sqrt{1} &   &   &  &\                \\
      & 0 & \sqrt{2} &   &  &    \\
      &   & 0 & \sqrt{3} &  &          \\
      &   &   & \ddots & \ddots  &
\end{array} \right)
\]
for the boson annihilator.

The idea is to represent $d_k$ and $d^\dagger_k$ by matrices which are
direct products of the $2 \times 2$ matrices
\ba\nonumber
I_{k}=\left(\begin{array}{cc}
      1 & 0   \\
      0 & 1
\end{array}\right)_{k}\hspace{3mm}&,&\hspace{5mm}
\sigma_{\!z\,k}=\left(\begin{array}{cc}
           1 & 0    \\
           0 & -1
\end{array}\right)_k   \nonumber \\
a_{k}=\left(\begin{array}{cc}
     0 & 1            \\
     0 & 0
\end{array}\right)_{k}\hspace{3mm}&,&\hspace{6.6mm}
a_{k}^{\dagger}=\left(\begin{array}{cc}
         0 & 0            \\
         1 & 0
\end{array}\right)_k
\ea
Here $k=1,2,3, \ldots \,$ is a label which distinguishes matrix factors
within each direct product. Consider then the infinite matrices
\ba\nonumber
d_{1}&=&a_{1}\times I_{2}\times I_3 \times \ldots              \nonumber  \\
d_{2}&=&\sigma_{\!z\,1}\times a_{2}\times I_{3}\times I_4 \times \ldots
                                                               \nonumber   \\
d_{3}&=&\sigma_{\!z\,1}\times \sigma_{\!z\,2}\times a_{3}\times I_{4}\times
                                    I_5 \times \ldots          \nonumber   \\
&\vdots&         \nonumber  \\
d_{k}&=&\sigma_{\!z\,1}\times\ldots\times\sigma_{\!z\,(k-1)}\times
    a_{k}\times I_{k+1}\times I_{k+2} \times \ldots          \label{d}
\ea
and their Hermitian adjoints:
\be
d_{k}^{\dagger}=\sigma_{\!z\,1}\times\ldots\times\sigma_{\!z\,(k-1)}\times
a_{k}^{\dagger}\times I_{k+1}\times I_{k+2} \times \ldots    \label{ddag}
\ee
Direct products have the property
\[  (A \times B).(C \times D) = (A.C) \times (B.D)   \]
where each dot indicates ordinary matrix multiplication, so the identities
\ba\nonumber
\{ a,\sigma_{\!z}\}_k\ = &0_k& =\ \{ a^\dagger ,\sigma_{\!z}\}_k\nonumber \\
\{ a , a^\dagger \}_k\ = &I_k& =\ \sigma^2_{\!z\,k}
\ea
in the $k^{\mbox{\small th}}$ factor space imply that (\ref{d}) and
(\ref{ddag}) form a matrix representation of (\ref{anticom}).

The next step is to convert these direct-product matrices to the operator
form (\ref{op}). To do that, we must choose a specific procedure for
ordering their rows and columns. (Different orderings are related by
similarity transformations.)

Let the rows and columns in the $k^{\mbox{\small th}}$ two-dimensional factor
space be labelled by the eigenvalues
\[  n_k = 0,1  \]
of the $k^{\mbox{\small th}}$ fermion number operator $d^\dagger_k d_k$.
Now treat the numbers $n_k$ as {\em binary digits\/} for the boson number
$n$ of Eq.\ (\ref{eigen}):
\be  n = \ldots n_k n_{k-1} \ldots n_2 n_1  \label{binary} \ee
Then the rows and columns of (\ref{d}) and (\ref{ddag}) can be ordered
according to increasing values $n = 0,1,2, \ldots$ of the boson number.
For example, the number 5 is 101 in binary notation, so the subspace labels
$n_1 = 1 = n_3$ with all other $n_k = 0$ correspond to the fifth row or
column.

To each integer such as the boson number $n$, we associate another integer
\be p(n) = \mbox{number of odd binary digits in $n$}
         = \sum_k n_k   \label{int} \ee
Thus $p(5)$ is 2, because $n_1$ and $n_3$ are the only binary digits of 5
which take the odd value 1.

Given these definitions, we can write the direct product of
$\sigma_{\!z}$ matrices in (\ref{d}) and (\ref{ddag}) as a
$2^{k-1} \times 2^{k-1}$ diagonal matrix
\be \sigma_{\!z\,1} \times \ldots \times \sigma_{\!z\,(k-1)}
       = \left( (-1)^{p(r)} \delta_{rs} \right)  \ee
where the integers $r$ and $s$ run from $0$ to $2^{k-1} - 1$. The effect of
the factor $a_k$ in (\ref{d}) is to lower all boson numbers $n$ with
$k^{\mbox{\small th}}$ binary digit $n_k = 1$ to the corresponding integers
with $n_k = 0$. Consequently, we obtain the operator expressions
\ba\nonumber
d_{1}&=&\sum_{m=0}^{\infty}\ket{2m}\bra{2m+1}      \nonumber  \\
d_{2}&=&\sum_{m=0}^{\infty}\Bigl( \ket{4m}\bra{4m+2}
                          -\ket{4m+1}\bra{4m+3}\Bigr) \nonumber \\
d_{3}&=&\sum_{m=0}^{\infty}\Bigl( \ket{8m}\bra{8m+4}-\ket{8m+1}\bra{8m+5}
        -\ket{8m+2}\bra{8m+6}+\ket{8m+3}\bra{8m+7}\Bigr)   \nonumber   \\
&\vdots&  \nonumber  \\
d_{k}&=&\sum_{m=0}^{\infty}\sum_{r=0}^{2^{k-1}-1} (-1)^{p(r)}
              \ket{2^{k}m+r}\bra{2^{k}m+2^{k-1}+r}    \label{dop}
\ea
and
\be
d_{k}^{\dagger} = \sum_{m=0}^{\infty}\sum_{r=0}^{2^{k-1}-1} (-1)^{p(r)}
               \ket{2^{k}m+2^{k-1}+r}\bra{2^{k}m+r}    \label{ddagop}
\ee

The final step is to rewrite (\ref{dop}) and (\ref{ddagop}) in terms of
the operators (\ref{boseops}) for a single Bose oscillator.

Within (\ref{dop}), let us replace the bra vector $\bra{2^{k}m+2^{k-1}+r}$
by $\bra{2^{k}m+r}$ and compensate by having the Bose annihilator $b$ act
$2^{k-1}$ times from the right:
\be
d_k = \left( (N+1)(N+2)\ldots (N+2^{k-1})\right)^{-\frac{1}{2}}
      \sum_m \sum_r (-1)^{p(r)}\ket{2^{k}m+r}\bra{2^{k}m+r}\, b^{2^{k-1}}
\label{diag} \ee
Notice that (\ref{diag}) involves a linear combination of projection
operators
\[
\sum^\infty_{m=0}\ket{2^{k}m+r}\bra{2^{k}m+r}
\]
Since these operators are diagonal, they must be functions of the Bose
number operator $N$ alone. The simplest example is the even-$n$ projector
\be
\sum^\infty_{m=0} \ket{2m}\bra{2m} = \cos^{2}(N\pi/2)  \label{Naka}
\ee
already considered by Naka \cite{Naka}. In our language, this projector
is associated with the identity
\be
\cos^{2}(n\pi/2) = 1 - n_1 = \left\{\begin{array}{ll}
                       1 \hspace{2mm} & \mbox{for $n$ even} \\
                       0 \hspace{2mm} & \mbox{for $n$ odd}
                        \end{array} \right.  \label{last}
\ee
where $n_1$ is the last binary digit of $n$ in (\ref{binary}).
The generalization of (\ref{last}) to the last $k$ binary digits of $n$
is
\ba
\prod^k_{\ell=1} \cos^{2}(n\pi/2^\ell)
&=& \prod^k_{\ell=1} \left( 1 - n_\ell \right) \nonumber \\
&=& \left\{\begin{array}{ll}
            1 \hspace{2mm} & \mbox{for $n$ an integer multiple of $2^k$} \\
            0 \hspace{2mm} & \mbox{otherwise}
             \end{array} \right.  \label{lastk}
\ea
This corresponds to the projection operator
\ba
{\cal P}_k(N)
&=& \sum^\infty_{m=0}\ket{2^{k}m}\bra{2^{k}m}  \nonumber  \\
&=& \cos^2(N\pi/2^k)\cos^2(N\pi/2^{k-1}) \ldots \cos^{2}(N\pi/2)
\label{projk} \ea
Then the projection operators in (\ref{diag}) can be obtained by shifting
the dependence of (\ref{projk}) on $N$ to $N-r$, for integers $r$ running
from $0$ to $2^{k-1}-1$:
\be
{\cal P}_k(N-r) = \sum^\infty_{m=0}\ket{2^{k}m+r}\bra{2^{k}m+r}
                = \prod^k_{\ell=1}\cos^{2}\left((N-r)\pi/2^\ell\right)
\label{proj}  \ee

In summary, the fermion annihilation operators $d_k$ can be constructed
from a Bose oscillator (\ref{boseops}) according to the prescription
\ba\nonumber
d_{1}&=&(N+1)^{-\frac{1}{2}}{\cal P}_1(N)\, b
            = (N+1)^{-\frac{1}{2}}\cos^{2}(N\pi/2)\, b  \nonumber  \\
d_{2}&=&\frac{1}{\sqrt{(N+1)(N+2)}}
            \Bigl({\cal P}_2(N) - {\cal P}_2(N-1)\Bigr)\, b^2  \nonumber  \\
d_{3}&=&\frac{1}{\sqrt{(N+1)(N+2)(N+3)(N+4)}}
  \Bigl({\cal P}_3(N) - {\cal P}_3(N-1) - {\cal P}_3(N-2) + {\cal P}_3(N-3)
                                       \Bigr)\, b^4   \nonumber    \\
&\vdots&
\ea
or generally
\be
d_{k} = \sqrt{N!/(N+2^{k-1})!}
      \sum_{r=0}^{2^{k-1}-1} (-1)^{p(r)} {\cal P}_{k}(N-r)\, b^{2^{k-1}}
\ee
where the integer $k$ runs from 1 to $\infty$, $p(r)$ is the
integer defined by (\ref{int}), and ${\cal P}_{k}(N-r)$ is the
projector (\ref{proj}). This completes the demonstration.

We do not anticipate that this procedure will supplant standard techniques
\cite{boson} for cases in which $k$ is a continuous parameter. Firstly, the
success of these techniques is hard to emulate. There is also the problem
that our method depends on the discreteness of the index $k$, and is not
easily adapted to the continuous case.

Instead, we are investigating the possibility that some variant of our scheme
could be used for numerical computations, possibly as a substitute for
putting fermions on the \vspace{3mm}lattice.

One of us (RJC) thanks members of the Institute for Theoretical Physics,
University of Berne, for their hospitality while this work was being
completed.


\begin{thebibliography}{10}
\bibitem{JW} P. Jordan and E. P. Wigner, Zeit. f\"{u}r Phys.
{\bf 47} (1928) 431;
P. Jordan, Zeit. f\"{u}r Phys. {\bf 93} (1935) 464;
M. Born and N. S. Nagendra Nath, Proc. Ind. Acad. Sci. {\bf 3} (1936) 318;
A. Sokolow, Phys. Z. der Sowj. {\bf 12} (1937) 148; O. Klein, J. Phys.
(USSR) {\bf 9} (1938) 1
\bibitem{Klaiber} B. Klaiber, in {\em Lectures in Theoretical Physics}, ed.
A. O. Barut and W. E. Britten (Gordon and Breach, New York, 1968), Vol.
{\bf X--A}, p. 141; J. H. Lowenstein and J. A. Swieca, Ann. Phys. (N.Y.)
{\bf 68} (1971) 172; G. F. Dell'Antonio, Y. Frishman, and D. Zwanziger,
Phys. Rev. D{\bf 6} (1972) 988
\bibitem{boson} S. Coleman, Phys. Rev. D{\bf 11} (1975) 2088; S.
Mandelstam, Phys. Rev. D{\bf 11} (1975) 3026; E. Witten, Commun. Math. Phys.
{\bf 92} (1984) 455
\bibitem{2d} Y. Frishman and J. Sonnenschein, Phys. Rep. C{\bf 223} (1993)
309; E. Abdalla, M. C. B. Abdalla, and K. D. Rothe,
{\em Non-perturbative Methods in Two-Dimensional Quantum Field Theory}
(World Scientific Publishing Co., 1991)
\bibitem{Naka} S. Naka, Prog. of Theor. Phys. {\bf 59} (1978) 2107
\bibitem{chinese} Tu-nan Ruan, Si-cong Jing, and An-min Wang, Europhys.
Lett. {\bf 23} (1993) 317
\end{thebibliography}
\end{document}